# Development of a bunching ionizer for TOF mass spectrometers with reduced resources

**Oya Kawashima**[*,1,2], **Satoshi Kasahara**[2], **Yoshifumi Saito**[1,2], **Masafumi Hirahara**[3], **Kazushi Asamura**[1], **and Shoichiro Yokota**[4]

[1] Institute of Space and Astronautical Science (ISAS), Japan Aerospace Exploration Agency (JAXA), 3-1-1 Yoshinodai, Sagamihara, Kanagawa 252-5210, Japan

[2] Department of Earth and Planetary Science, School of Science, The University of Tokyo, 7-3-1 Hongo, Bunkyo-ku, Tokyo 113-0033, Japan

[3] Institute for Space-Earth Environmental Research (ISEE), Nagoya University, Furo-cho, Chikusa-ku, Nagoya, Aichi 464-8601, Japan

[4] Department of Earth and Space Science, School of Science, Osaka University, 1-1 Machikaneyama-cho, Toyonaka-shi, Osaka 560-0043, Japan

* Corresponding author.  E-mail address: kawashima.oya@jaxa.jp (Oya Kawashima)

Abstract: In some types of mass spectrometers, such as Time of Flight mass spectrometers (TOF-MSs), it is necessary to control pulsed beams of ions. This can be easily accomplished by applying a pulsed voltage to the pusher electrode while the ionizer is continuously flowing ions. This method is preferred for its simplicity, although the ion utilization efficiency is not optimized. Here we employed another pulse-control method with a higher ion utilization rate, which is to bunch ions and kick them out instead of letting them stream. The benefit of this method is that higher sensitivity can be achieved; since the start of new ions cannot be allowed during TOF separation, it is highly advantageous to bunch ions that would otherwise be unusable. In this study, we used analytical and numerical methods to design a new bunching ionizer with reduced resources, adopting the principle of electrostatic ion beam trap. The test model experimentally demonstrated the bunching performance with respect to sample gas density and ion bunching time using gas samples and electron impact ionization. We also conducted an experiment connecting the newly developed bunching ionizer with a miniature TOF-MS. As a result, the sensitivity was improved by an order of magnitude compared to the case using a non-bunching ionizer. Since the device is capable of bunching ions with low voltage and power consumption, it will be possible to find applications in portable mass spectrometer with reduced resources.

# 1. Introduction

A mass spectrometer (hereafter MS) can be divided into several basic parts including an ionizer, a mass analyzer, and a detector. Among them, an ionizer directly affects the performance of MS, and thus have been well designed for laboratory and commercial use. As well as improving the efficiency to generate ions, efforts have been made to optimize the performance of ion optics to control the ion movement. For example, Einzel lenses, quadrupole ion lenses, and multipole-rods or ion funnels have been adopted to maximize ion transmission efficiency to mass analyzers[1].

Types of MSs range from quadrupole, Time of Flight (hereafter TOF), magnetic, Fourier transform ion trap, RF ion trap, electrostatic ion trap, and other relatively minor ones. Electrostatic ion trap, the scope of this study, uses an electrostatic field instead of a RF electric field to trap ions in a limited space. There are two main methods of electrostatic ion traps that have been established so far: electron beam ion trap (hereafter EBIT) and electrostatic ion beam trap (hereafter EIBT). EBIT is the method that uses electron beams to create local minima of potential and trap ions there with small kinetic energies. As known from Earnshaw's theorem, the Laplace equation can only give solutions with local potential extrema in the vicinity of the space charge such as free electrons. As an example of its application, EBIT has been used to study highly charged ions by using electron beams to trap ions and at the same time remove multiple electrons from the trapped ions[2]. EIBT, on the other hand, is a method to trap ions with a certain high kinetic energy. The dynamics of the ion beam is controlled under electrostatic fields, and thus trappable ions are characterized by the energies. Examples are found in Orbitrap$^{TM}$ which uses rotational and bouncing motion of ions[3], and electrostatic linear ion trap which uses reflection motion of ions between two ion mirrors[4–11].

In this study, we designed EIBT as an ionizer and combined it with a mass spectrometer that requires pulsed ion current, such as TOF-MS, to improve sensitivity (Figure 1). Classically, the pulsed ion current is obtained by applying a pulsed voltage to a pusher electrode while an ion source is continuously streaming ions. This is essentially a loss-making operation in terms of the ion utilization. The ion inlet shutter must be closed until the end of the mass separation in order to preserve information on when the ions started. Any ions that are streamed during this period are lost by colliding with the shutter. On the other hand, there is another popular pulse control method occasionally used in laboratory high-end equipment, where ions are bunched and kicked out instead of being streamed. The benefit of this method is that higher sensitivity can be achieved by bunching ions that would otherwise be unusable. With flexibility in controlling the amount of ions in bunching, saturation or reaching upper limit of quantitation can be avoided, making operation equivalent to that of a streaming ionizer possible.

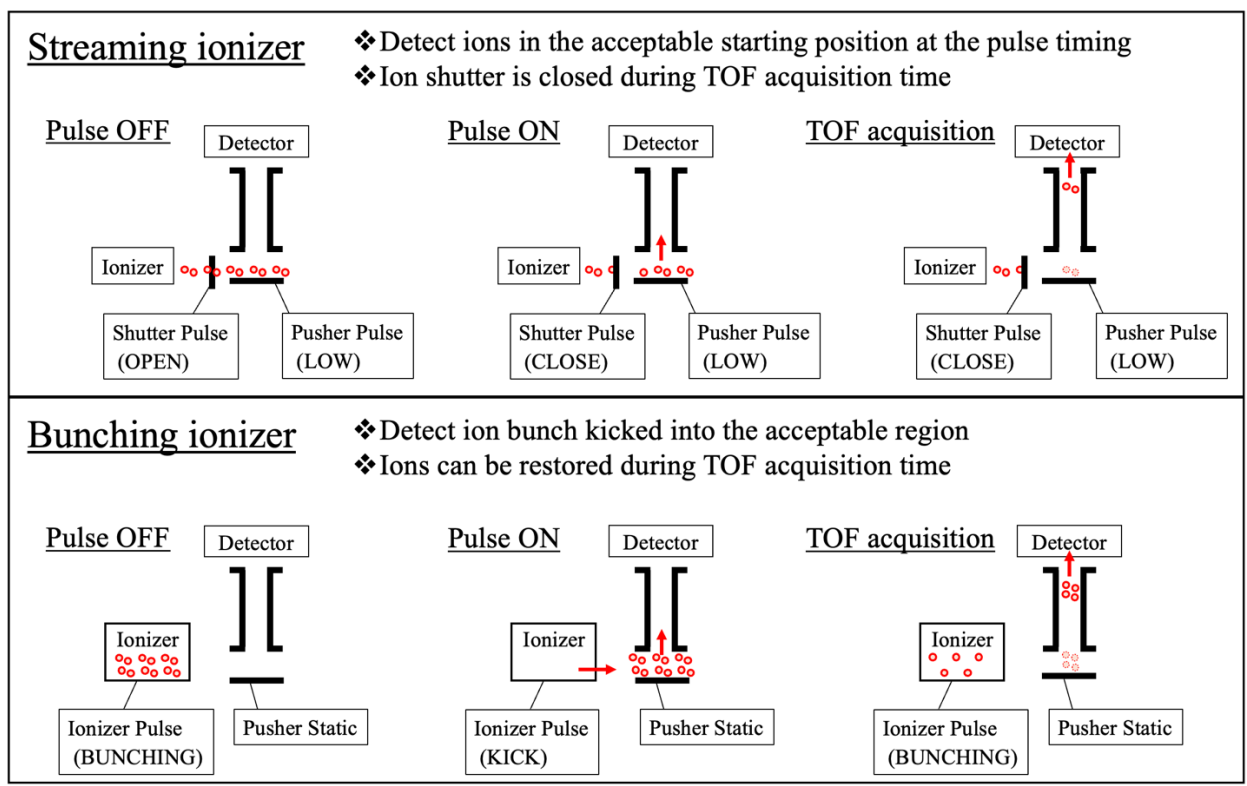

**Figure 1.** Schematic diagram of TOF with streaming ionizers (upper) and bunching ionizers (bottom).

In general, MS performance is often a trade-off between the resources (e.g., size of the ion optics and power consumption), and therefore the sensitivity issue becomes more apparent in the development of miniaturized MSs. There are some situations where performance should be compromised for resources. The recent increase in demand for portable MSs for environmental monitoring, for volcanic gas measurements on aircraft, and for analyses of extraterrestrial samples on spacecraft is illustrative. In such small MSs, it is better to have ionizers that compensate for the reduced sensitivity. Previous publication reported that the ROSINA RTOF on board the ESA's Rosetta spacecraft employed a type of EBIT that utilizes the slight potential drop due to the electron beam potential so as to retain the generated ions for a certain time[12]. When using electron impact ionization (hereafter EI), it is simplest to utilize the accompanying electron beam potential, but in order to provide a considerable potential drop such as larger than ~1 mV, the electron beam density must be significantly high ($>\sim$1 A mm$^{-2}$)[13, 14]. The electron beam density used in ordinary EI is ~1 mA mm$^{-2}$, which may threaten to provide insufficient potential drop. Actually, the miniature EBIT in a laboratory equipment uses specific beam-focusing optics for electrons to obtain adequate potential drop[15]. In addition to EBIT, it is possible to consider bunching ionizers adopting other ion trap principles. In fact, previous high-performance instruments are utilizing RF ion traps to design ion bunchers. Examples of RF ion bunchers can be found in LCMS-IT-TOF$^{TM}$ [16], C trap[17] and quadrupole ion trap[18, 19]. In these instruments, by allocating additional electrical resources including a RF power supply (~ kV amplitude and ~ MHz frequency) that generally requires 5–10 W[20], not only pulsed ion currents but also MS$^n$ tandem analysis are enabled. This must be the best choice if there is room for the additional resources, but as evidenced by the lack of this option on previous small MSs, it is worthwhile to study a rather simplified buncher. Here, EIBT is a more feasible and achievable option for the future MSs with reduced resources due to the lower power consumption required to trap ions and the fact that the high-TRL pulse voltage generator of streaming ionizers can be applied directly. Note that, although

EIBT has been used as MS in previous studies, there have been no reports that have verified the performance of EIBT as a bunching ionizer. In this study, we report that our new EIBT-based miniature bunching ionizer successfully worked with low power consumption (~2.2 W at 1 kHz operation, see section 3) and contributed to the sensitivity enhancement of a small TOF-MS. In section 2, we revisited the EIBT ion-trap condition to design the bunching ionizer. Section 3 presents the results of experimental performance tests, including bunching tests by the ionizer itself and sensitivity tests in combination with a small MS. Section 4 summarizes the results of the development and discusses the general applicability to the future instruments.

## 2. Design Description

### 2.1. Analytical expression

EIBT is a device invented to store ion beams with kinetic energy using the electrostatic field. The trap can be achieved by positioning ion mirrors oppositely just as in laser cavities which keep photon beams reflecting within mirrors. The EIBT has been utilized in laboratory experiments with energetic ions such as measurements of ion lifetime and ion cooling efficiency[4, 5], and in miniaturization of mass analyzer[8, 9].

In this study, we revisited the design of the EIBT for application to bunching ionizers. The schematic image of the device is shown in Figure 2. The trap consists of two ion mirrors and two ion lenses, all of which are combinations of disk-shaped electrodes with a hole in the middle. The outermost ion mirrors are in the role of reflecting the ion beam to trap it inside. If the ion mirror diverges the ion beam as shown in the upper left of Figure 2 (i.e., acts as a convex mirror), ion lenses known as Einzel lenses should be used to cancel the beam divergence.

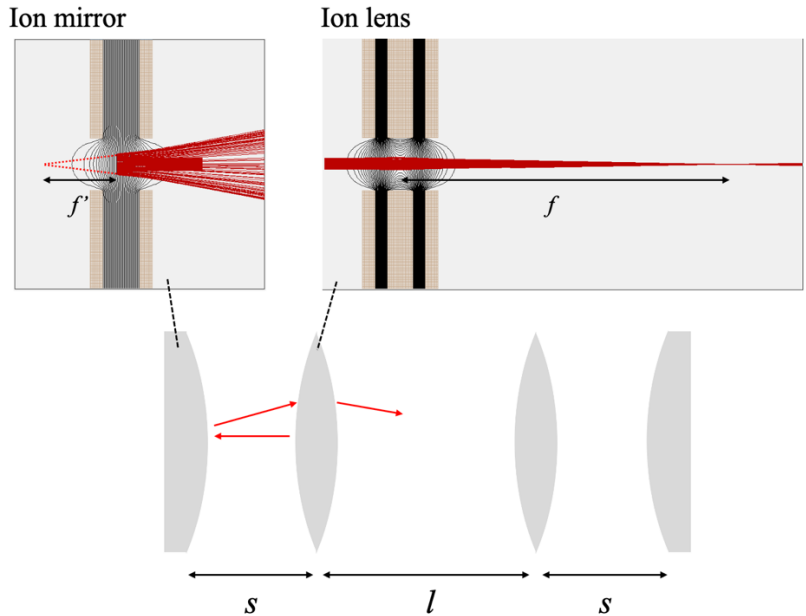

**Figure 2.** Schematic of the EIBT with examples of ion beam trajectories entering ion mirrors and ion lenses. In the simulation example of the mirror and lens, ion beam trajectories are shown in red, equipotential lines are shown in black. Brown rectangles are cross sections of disk-shaped electrodes.

When designing a device as a combination of optical elements, ray transfer matrices are beneficial. As an analytical expression, each optical component, including free space, is described by a 2×2 ray transfer matrix that converts incoming ion-beam rays $(r, \theta)$ into outgoing rays $(r', \theta')$. For example, an ion mirror matrix can be written as:

$$\begin{pmatrix} r' \\ \theta' \end{pmatrix} = \begin{pmatrix} 1 & 0 \\ -\frac{1}{f'} & 1 \end{pmatrix} \begin{pmatrix} r \\ \theta \end{pmatrix} \quad (1)$$

where $f'$ is the focal length of the mirror. Similarly, an ion lens matrix can be written as:

$$\begin{pmatrix} r' \\ \theta' \end{pmatrix} = \begin{pmatrix} 1 & 0 \\ -\frac{1}{f} & 1 \end{pmatrix} \begin{pmatrix} r \\ \theta \end{pmatrix} \quad (2)$$

where $f$ is the focal length of the lens. In addition, if we consider the field-free region of length $d$ to be an optical element, the beam travels with an initial inclination, thus:

$$\begin{pmatrix} r' \\ \theta' \end{pmatrix} = \begin{pmatrix} 1 & d \\ 0 & 1 \end{pmatrix} \begin{pmatrix} r \\ \theta \end{pmatrix} \quad (3)$$

Considering equations (1)-(3), the transfer matrix for a half round trip in the trap, which is the minimum unit of repeated components when considering an infinite trap, is as follows:

$$\begin{pmatrix} 1 & 0 \\ -\frac{1}{f'} & 1 \end{pmatrix} \begin{pmatrix} 1 & s \\ 0 & 1 \end{pmatrix} \begin{pmatrix} 1 & 0 \\ -\frac{1}{f} & 1 \end{pmatrix} \begin{pmatrix} 1 & l \\ 0 & 1 \end{pmatrix} \begin{pmatrix} 1 & 0 \\ -\frac{1}{f} & 1 \end{pmatrix} \begin{pmatrix} 1 & s \\ 0 & 1 \end{pmatrix}$$

$$= \begin{pmatrix} \left(1-\frac{l}{f}\right)\left[\left(1-\frac{s}{f}\right)\left(1-\frac{s}{f'}\right)-\frac{s}{f'}\right]-\frac{l}{f'}\left(1-\frac{s}{f}\right)-\frac{l}{f} & \left(1-\frac{l}{f}\right)\left(2-\frac{s}{f}\right)s-\frac{ls}{f}+l \\ -\frac{1}{f}\left[\left(1-\frac{s}{f}\right)\left(1-\frac{s}{f'}\right)-\frac{s}{f'}\right]-\frac{1}{f'}\left(1-\frac{s}{f}\right)-\frac{1}{f} & -\frac{1}{f}\left(2-\frac{s}{f}\right)s-\frac{s}{f'}+1 \end{pmatrix} \equiv K = \begin{pmatrix} A' & B' \\ C' & D' \end{pmatrix} \quad (4)$$

where the distance between the lens and the mirror is $s$, and the distance between two lenses is $l$. Here, a necessary condition for an infinite ion trap is equal to the condition that $(r, \theta)$ do not diverge after repeated manipulation of the matrix $K$. Translating this condition as an eigenvalue problem, we consider the following equation about eigenvalues $\lambda$ of the system:

$$K \begin{pmatrix} r_0 \\ \theta_0 \end{pmatrix} = \lambda \begin{pmatrix} r_0 \\ \theta_0 \end{pmatrix} \quad (5)$$

By transforming the equation (5) using the unit matrix $E$, we obtain:

$$[K - \lambda E] \begin{pmatrix} r_0 \\ \theta_0 \end{pmatrix} = 0 \quad (6)$$

The determinant of each ray transfer matrix is 1, and therefore:

$$\det(K) = A' D' - B' C' = 1 \quad (7)$$

The eigenequation of the system is as follows:

$$\lambda^2 - \left(A' + D'\right)\lambda + 1 = 0 \quad (8)$$

Using the expression of tr($K$)= $A'+D'$, the eigenvalues are expressed as follows:

$$\lambda_\pm = \frac{\text{tr}(K)}{2} \pm \sqrt{\left(\frac{\text{tr}(K)}{2}\right)^2 - 1} \quad (9)$$

If tr($K$)/2 is greater than 1, there should be an eigenvalue greater than 1. This means that vectors ($r$, $\theta$) are stretched in the corresponding eigendirection by the repeated manipulation of matrix $K$, which causes the ions to collide with the electrodes. Therefore, the stability condition for an infinite trap is that the magnitude of both eigenvalues is less than 1, which can be expressed as:

$$\left|\frac{\text{tr}(K)}{2}\right| \leq 1 \tag{10}$$

$$\therefore 0 \leq \left(1-\frac{s}{f}\right)\left(1-\frac{l}{2f}\right) - \frac{1}{4f'}\left(1-\frac{s}{f}\right)\left(2s+l-\frac{ls}{f}\right) \leq 1 \tag{11}$$

Inequality (11) shows that the ion beam trap condition can be satisfied by choosing the appropriate four parameters ($s$, $l$, $f$, $f'$). We chose $s$ = 10 mm and $l$ = 5 mm by setting the minimum electrode thickness at 1 mm for the manufacturing reason, and setting the minimum distance between electrodes at 1 mm for reducing discharge risks. By fixing these values, (11) is reduced to two variables, and the ($f$, $f'$) region satisfying the inequality is represented by the shaded region in Figure 3. As noted above, an ion mirror which is a combination of simple disk-shaped electrodes act as a convex mirror and thus the choices of parameters are limited to the red shaded regions. Of these red regions, we chose the smaller $1/f$ and $1/f'$ which can be achieved at the lower voltages. Finally, the parameters ($s$, $l$, $1/f$, $1/f'$) = (10, 5, 0.05, -0.2) or ($s$, $l$, $f$, $f'$) = (10, 5, 20, -5) were selected for consistency of geometrical size and focal length.

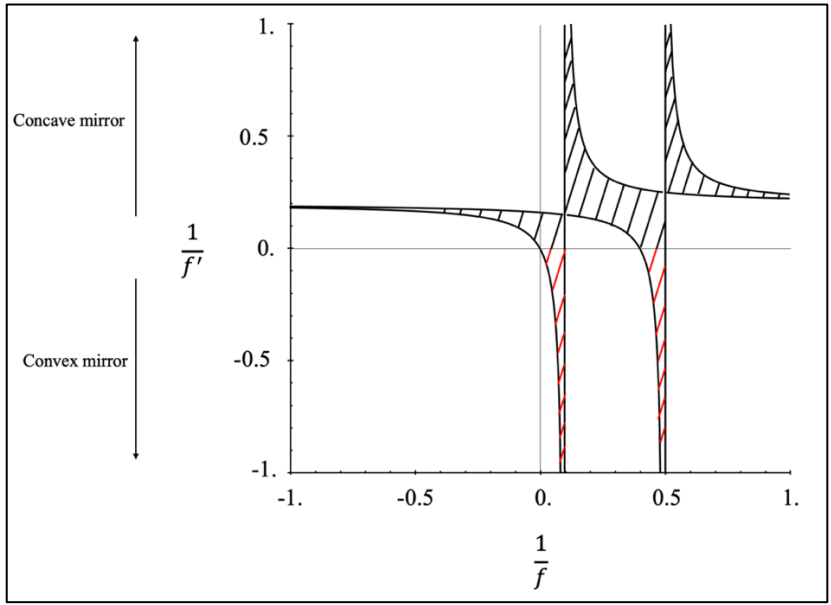

**Figure 3.** Parameter region satisfying inequality (11) when $s$ = 10 mm and $l$ = 5 mm. The horizontal axis is the inverse of the lens focal length, and the vertical axis is the inverse of the mirror focal length.

## 2.2. Numerical simulation

In the following, we present the design of the EIBT optics satisfying ($s$, $l$, $f$, $f'$) = (10, 5, 20, -5). While $s$ and $l$ are parameters solely depend on geometry, $f$ and $f'$ depend on several variables including the geometry, the voltages of the lens and mirror, and the energy of the ion beam. To fix the geometry of the device, it is therefore necessary to determine the voltages to be used and the energy of the ion beams to be trapped.

To consider the voltages and ion beam energies, we first determined the ion trapping method. When obtaining ions inside traps, there are two choices: an external method, in which the electric potential is temporarily opened at the moment of the ion-beam entry and is immediately returned to the trapping status, or an internal method, in which ions are generated in the trappable region and are subsequently trapped. For our purposes, it is reasonable to choose the internal method to bunch ions while ionizing neutral gases. This means we needed to optimize the ionization efficiency at the trappable region. We employed EI as the ionization method, and to maximize ionization efficiency using EI, electrons with kinetic energies of 60–80 eV should be bombarded with the neutral gases[21]. Thus, we set the potential difference between the electron gun and the trappable region to 60–80 V.

Assuming a typical ion beam energy of 70 eV, numerical simulation is useful to determine the mirror and lens voltages that satisfy $(f, f') = (20, -5)$. Using SIMION® 8.0 software, we confirmed that a mirror with a combination of 80 V, 57 V, and 0 V electrodes and a lens with a combination of 0 V and 30 V electrodes satisfy these conditions. Figure 4 shows the summary of the calculations integrating these electrodes as one device. Figure 4a shows the electron beam trajectories with equipotential lines and electrodes in the new device. We employed a thermionic electron gun combined with Whenelt-type electrodes capable of ~10 mA mm$^{-2}$ sr$^{-1}$ brightness at 80 V[22]. Note that space-charge effects (i.e., contribution of point charges to electric potential) were ignored in this simulation to reduce the calculation resources. It is possible that the electron beam may become more dispersed depending on the beam density. Furthermore, we cannot rule out the possibility that the space-charge effect of the electron beam could distort the ion trapping potential and make trapping impossible; however, we report in the following section that, at least under the experimental conditions we used, there seemed to be no such problems. Figure 4b shows the example of ion distribution during bunching. In this calculation, the ions are assumed to be randomly produced in the EIBT, and those that have proper initial conditions are trapped. Figure 4c shows the desired initial position of ions to be successfully trapped. Ions acquire different potential energies and initiate subsequent capture motions depending on where they are generated, thus the capture efficiency differs between the area. The colored areas on the heat map have high trapping efficiency, and the ions generated at these areas accumulate in the EIBT.

Similar ion optics to our system can be found in electrostatic ion beam trap[10] and Autoresonant ion trap[8]. Figure 5 shows a comparison of the design and stable regions. Our design uses a similar concept as model 2 of ion-trap resonator[10], but a comparison of the stability regions shows that wider $1/f'$ and $1/f$ are acceptable. This means that the previous ion-trap resonator is designed to stably trap limited energies, making it suitable for energetic ion experiments. Instead, our new design allows for the wider range of parameters that affect the focal length of the lens or mirror, including ion initial positions and energies. In the Autoresonant ion traps[8], two concave ion mirrors are used but no ion mirrors. This corresponds to setting the parameters in inequality (11) as $1/f=0$ and $l=0$, thus the stability condition is $s/2 <= f$. This design is also useful for miniaturization, as seen in a commercial portable MS (e.g.,Granville-Phillips Series 830 Vacuum Quality Monitor); in this study, we employed ion lenses to increase the number of control electrodes for ion kicked out (see below).

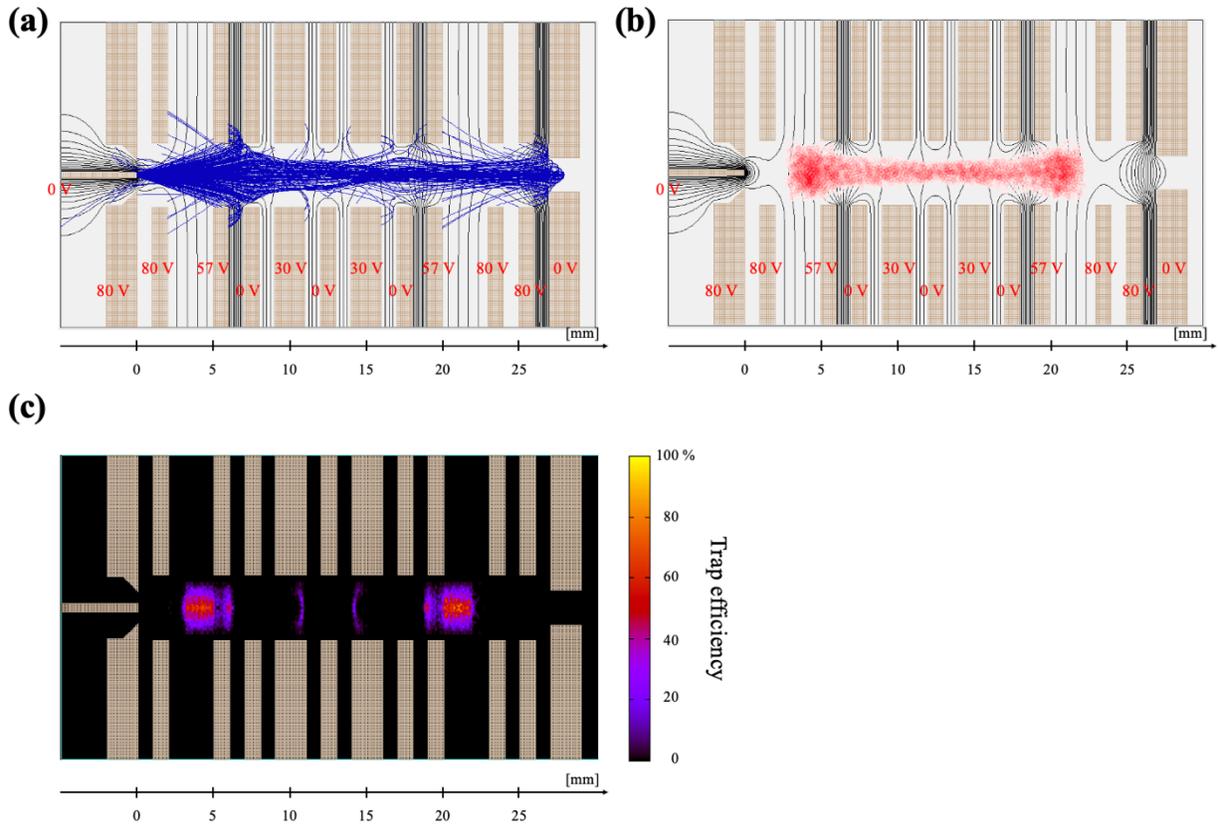

**Figure 4.** (a) Electron beam trajectories (blue lines). Electrons emitted from a 0 V electron gun (left center) are accelerated and injected into the EIBT, ionizing background neutral gas. (b) An example of spatial distribution of trapped ions (red points). In both figures, the black lines show equipotential lines in 10 V increments, the brown rectangles show cross sections of the disk-shaped electrodes. The red numbers are the voltages applied to the electrodes. (c) A heatmap of trap efficiency showing how much fraction of ions produced at a given area are subsequently trapped.

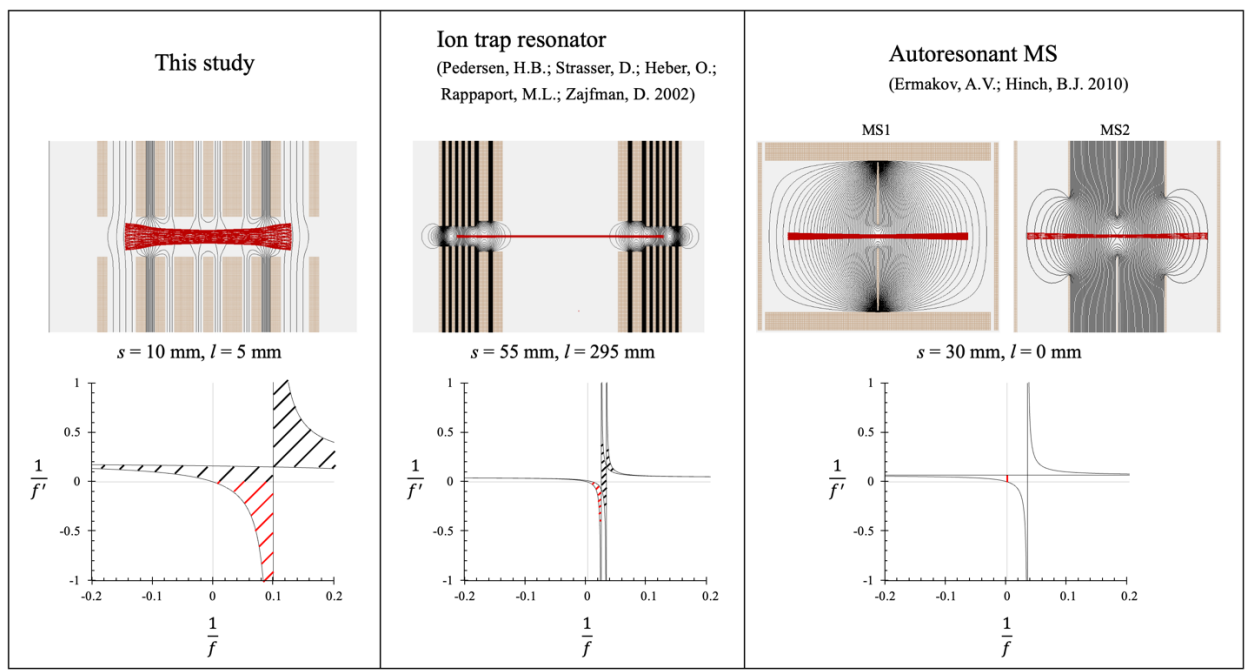

**Figure 5.** Comparison of ion trajectories and stability conditions with other similar EIBT systems. Ion trajectories are calculated using SIMION 8.0, referring to the length and voltage parameters shown in the publications. The area of the stability conditions adopted in the EIBT design are shaded in red. (Left column) The design of bunching ionizer in this study. (Center column) The design of ion trap resonator[10]. (Right column) The design of Autoresonant MS1 and MS2[8].

The above results confirm that the generated ions should accumulate successively in the EIBT. As the next step, it is equally important to verify whether the bunched ions can be kicked out toward mass analyzer. By switching the voltage from the configuration shown in Figure 4, bunched ions should be kicked out so that the timing of the TOF start can be synchronized. Figure 6 shows the calculation results simulating the kick-out, where voltage switching is triggered at $t = 10$ μs. Note that these calculations are for $N_2^+$ ions ($m/z = 28$), and different mass-to-charge ratios will change the TOF and other results. The voltage switching requires a finite time that may not be negligible; in this study, with the help of EOSTECH Ltd., Kanagawa, Japan, we used a proprietary fast switcher with a time constant of less than ~10 ns. Therefore, the calculation assumes the voltage waveform switches linearly from low to high between 10.00 to 10.01 μs. An example of ion trajectories after 10.01 μs is shown in Figure 6a. The voltages after switch are determined such that the ions are kicked out forward. Since the maximum energy of the trapped ions is 80 eV, the potential wall on the side of electron gun is set to be 80 V. On the side close to the exit, the 20 V electrode acts as an Einzel Lens to focus the beam. The potentials of the other electrodes were optimized through SIMION calculations to kick out the maximum amount of ions while minimizing beam divergence. In addition, the Wehnelt voltage is switched to -100 V to suppress the emission from electron gun, since continuous ion generation after switching is undesirable as a pulsed ion current source. Figure 6b shows a heatmap about the number of ions passing through the "Profile Plane" in Figure 6a. The distribution of ions is focused on the center owing to the lensing effect set up at the end. Figure 6c shows a heatmap about the TOF and kinetic energy of ions that reached at the "Profile Plane". The largest peak appeared at ~10.5 μs and ~75 eV, corresponding to the ejection of ions that were $z = 6–8$ mm at the moment of the switch, where $z$ is the axial position as shown in Figure 6a. The ions are trapped with energies corresponding to the potential at the location where they are generated, so it is reasonable that ions are distributed within 60–80 eV. The distribution of energies, however, is not desirable as this can lead to poor performance of the MS. In order to obtain a monoenergetic ion beam, it is required to further improve the design such as by focusing the electron beam on a certain narrow region. For example, an improved design would focus the electron beam in a high-efficiency region (Figure 4c) perpendicular to the EIBT axis. Note that this vertical electron beam method has been actually implemented in autoresonant ion trap[8, 9], showing that this option is effective. We also admit that among the ions that are trapped, populations that are far out of time and energy as a result of kicking out cannot be used because they may significantly degrade the performance of the MS. Although the current design has limitations in this regard, but it may be possible to open a slit perpendicular to the axis of the EIBT and extract the ion, such as used in linear ion traps[23].

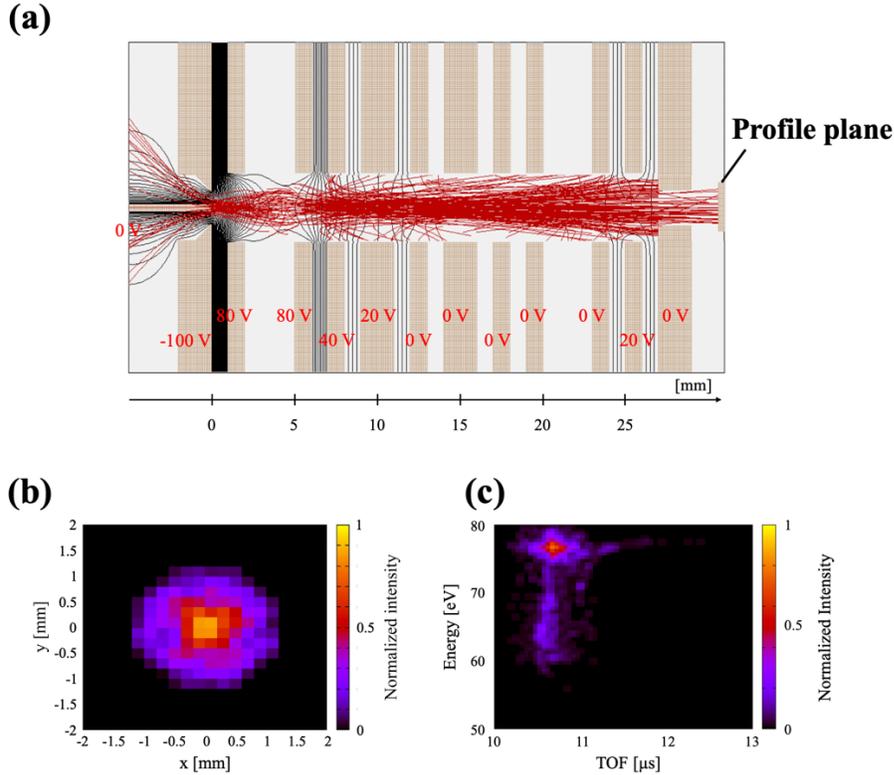

**Figure 6.** Simulation of ion bunch kickout. Ions are assumed to be $N_2^+$ ($m/z$ = 28). (a) Ion trajectories (red lines) after voltage switching. The black lines show equipotential lines in 10 V increments, and the brown rectangles show cross sections of the disk-shaped electrodes. (b) A heatmap about the number of ions reaching the "Profile Plane" in Figure 6a. (c) A heatmap about TOF and energy of ions reaching the "Profile Plane" in Figure 6a.

## 3. Experimental results and discussion

### 3.1. Demonstration of ion bunching

Here we present the experimental results using a test model fabricated according to the design in Section 2. The experimental operation to demonstrate the bunching capability is shown in Figure 7. The voltage was switched between bunching mode (Mode I) and kick-out mode (Mode II). The duration of Mode I ($T_1$) was variable, while the duration of Mode II ($T_2$) was controlled to $T_2$=1 ms-$T_1$ when $T_1$<100 μs, and fixed to $T_2$=1 ms when $T_1$<100 μs. In bunching mode, ions were continuously generated at the trappable region by EI, and subsequently trapped. As the analyte sample gas, pressure-regulated 99.9% dry $N_2$ was used. In the kick-out mode, the ion bunch was ejected, and the electron beam was suppressed by switching the Wehnelt electrode to -100 V. As noted above, voltage switching between these two modes was performed by a proprietary high-speed switcher with a time constant of less than ~10 ns. The power consumption was stable at ~0.8 W during Mode I (~0.7 W for electron gun and ~0.1 W for electrodes), and additional power is required for Mode II to switch voltages. At a kick-out frequency of 1 kHz, the total power consumption of the bunching ionizer for Modes I and II totaled ~2.2 W. Note that the power consumption increases with the frequency of switching; in this experiment, ~2.2 W was the maximum value because we only used switching frequencies lower than 1 kHz. Considering that RF power sources for ion guides and ion traps require generally 5–10 W[20], the new bunching ionizer is consuming lower power. As a detector of the ion bunch, we used an inverting charge-sensitive amplifier, which outputs a opposite polarity voltage proportional to the total amount of ions injected. The charge-sensitive amplifier uses a feedback resistor $R_f$ and a feedback capacitor $C_f$ and returns a voltage pulse characterized by a decay time expressed as $\tau = R_f C_f \sim 30$ μs. Therefore, the detector voltage at the timing of the ion bunch injection was negative and appeared as the transient waveform.

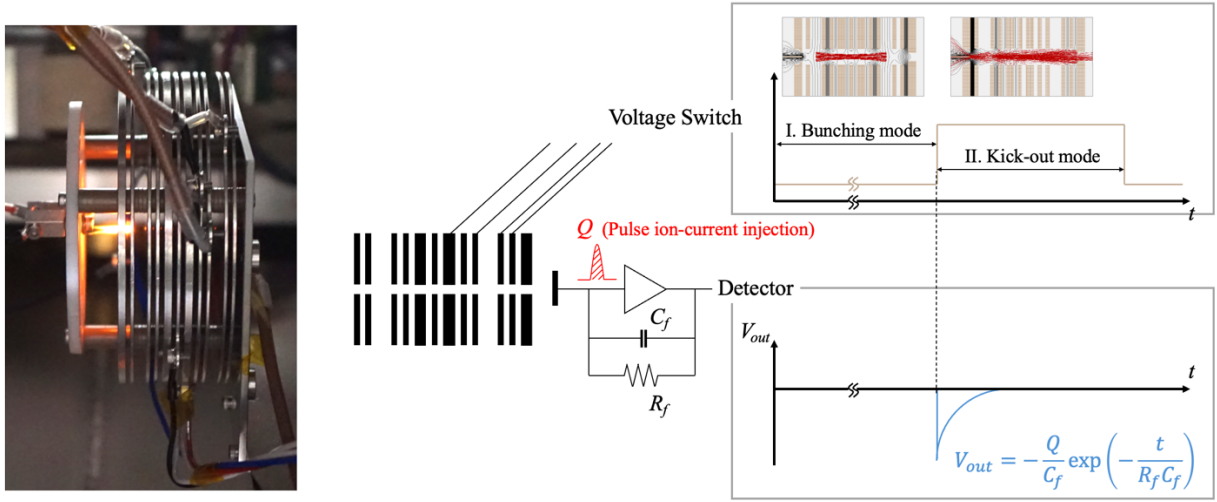

**Figure 7.** The concept of experimental operation to demonstrate the bunching capability.

In Figure 8, we present an example of experimental data showing the signal of ion bunch. Figures 8a and 8b show the time series data of the detector output when Mode I is maintained for 199 ms, with a different range on the horizontal and vertical axes. On the detector output, sharp noises due to voltage switching between Mode I and Mode II appeared at the timing of 0 and 1000 μs. We subtracted blanks from the experimental data to cancel this sharp noise, but because the level of noise varies slightly from trial to trial, then the sharp noise is still noticeable in some data. Other notable findings are that the pressure-dependent signal appeared at the timing of the switch from Mode I to Mode II, corresponding to the positive ion bunch incidence. The pressure dependency can be seen from the different amplitude of three colored lines indicating background $N_2$ pressure during the experiment. This result supports the basic understanding of EI, in that higher gas pressure produces more ions and therefore higher ion bunch currents.

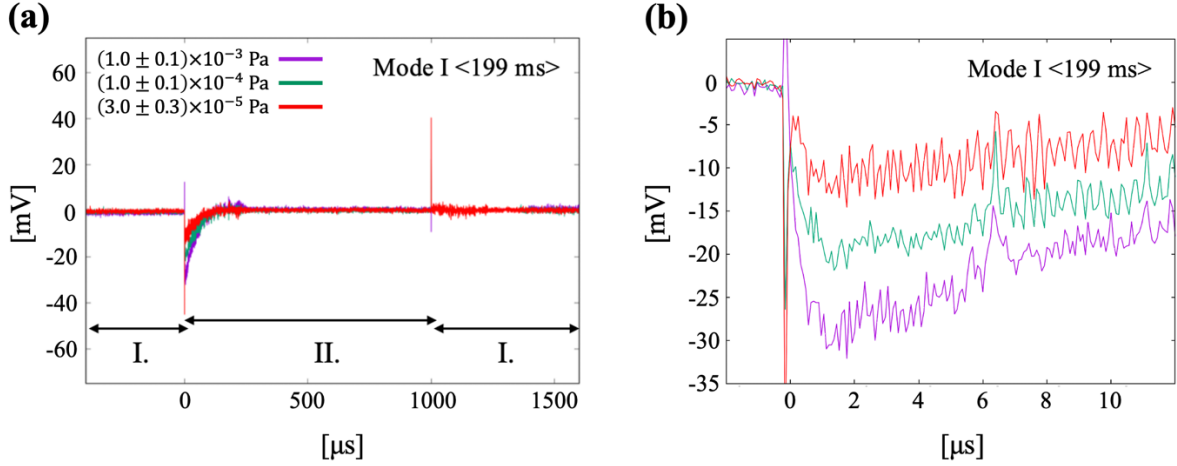

**Figure 8.** An example of experimental data showing the signal of ion bunch. (a) Detector output obtained when the duration of Mode I is 199 ms. The horizontal axis shows the time, and the vertical axis shows the output voltage of the detector. Three different colored lines indicate the different pressure of N₂ gas at which the experiment was conducted, as shown in the legend. (b) Same data as in Figure 7a, but zoomed in from 0 to 10 μs on the horizontal axis and from -35 to 0 mV on the vertical axis.

Figure 9 lists the results of the data with the duration of Mode I as a variable. As a rough trend, the signal intensity increases as the Mode I time increases, which is normal given that the more ion generation time, the more ions are generated. Figure 10 summarizes these data while converting the voltages to the number of injected ions based on the calibration from oscilloscope test signals (1390 # mV⁻¹). Here we can identify two phases: one in which ion abundance increases as the duration of Mode I increases (<~10⁵ μs), and one in which saturation occurs (>~10⁵ μs). The saturation can be explained by the balance between the ion production rate and the ion loss rate due to collisions with neutral particles. The following equation theoretically expresses the change in the trapped ion amount $N_i$ per unit time:

$$\frac{dN_i}{dt} = n_g n_e \sigma_{EI} \bar{v}_e - N_i \sigma_{loss} \bar{v}_i \tag{12}$$

where $n_g$ is the number density of neutral gases, $n_e$ is the number of electrons, $\sigma_{EI}$ is the ionization cross section of EI method, $\bar{v}_e$ is the mean velocity of electrons, $\sigma_{loss}$ is the cross section of ion loss per unit volume, and $\bar{v}_i$ is the mean velocity of trapped ions. Solving the equation (12) yields:

$$\rightarrow N_i = \frac{n_g n_e \sigma_{EI} \bar{v}_e}{\sigma_{loss} \bar{v}_i}[1 - \exp(-\sigma_{loss}\bar{v}_i t)] \tag{13}$$

When saturation occurs (i.e., $t \rightarrow \infty$), $N_i$ converges toward certain value depending on $n_g$. Before saturation occurs, $N_i$ is limited depending on $t$. In both cases, a factor which is the function of $n_g$ (i.e., $n_g n_e \sigma_{EI}\bar{v}_e/\sigma_{loss}\bar{v}_i$) controls $N_i$. At first glance the factor appears to be a linear function of $n_g$, but $\sigma_{loss}$ is also a function of $n_g$ as it includes complicated effects such as collision frequency and the degree of momentum exchange, resulting in a nonlinear relationship between $N_i$ and $n_g$. To understand the results better, it is also helpful to estimate the saturation timescale. Focusing on the order estimation, assume that $\bar{v}_i$ of trapped $N_2^+$ is at the order of $10^6$ cm s⁻¹. It is difficult to determine the loss cross section; based on publications[24, 25] that compiled experimental data of cross sections for collisions between $N_2$ and $N_2^+$ and those for collisions between electron and $N_2$, $N_2^*$, and $N_2^+$, the cross sections of elastic scattering (~5×10⁻¹⁵ cm²) appears to be ~5 times higher than that of charge transfer (~1×10⁻¹⁵ cm²), and more than 10 times higher than that of electron capture and fragmentation ($<1\times10^{-16}$ cm$^2$). If we substitute the loss cross section of $N_2^+$ as $\sigma_{loss} \sim 10^{-15}$ cm$^2$ for representative and rewrite $n_g$ as $n_g \sim p \times 10^{14}$ cm$^{-3}$, where $p$ is the pressure in pascals, the exponential part of the equation (13) can be rewritten as follows:

$$N_i = \frac{n_g n_e \sigma_{EI} \bar{v}_e}{\sigma_{loss} \bar{v}_i}[1 - \exp(-0.1pt)] \qquad (14)$$

where $t$ is the duration of Mode I in microseconds. This means that, at a pressure of $p \sim 10^{-3}$ Pa, the timescale for saturation would be $10^4$–$10^5$ μs which seems to adequately explain the experimental results in Figure 10. Note that the saturation timescale estimation may change depending on the ion loss cross section and the background pressure. As more detailed estimation, the results of quantitative simulations incorporating the SIMION 8.0 elastic hard sphere collision model are shown in Figure 9 by lines, which indicate the saturation timescale of $10^4$–$10^6$ μs depending on the pressure. In this calculation, the collision cross section between $N_2^+$ and $N_2$ was set to $3.0 \times 10^{-15}$ cm$^2$ to best match the experimental results.

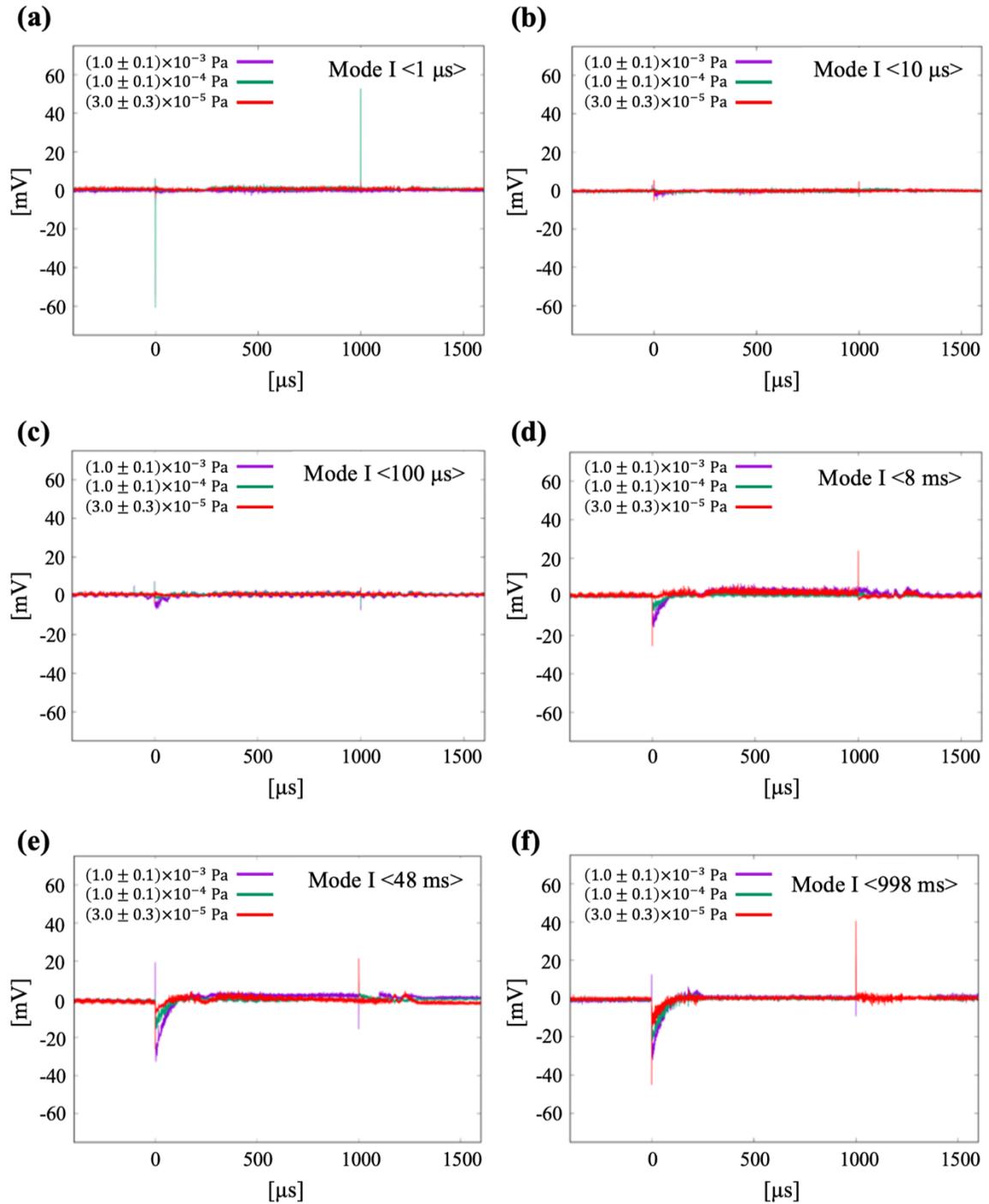

**Figure 9.** Detector output obtained when the duration of Mode I is (a) 1 μs, (b) 10 μs, (c) 100 μs, (d) 8 ms, (e) 48 ms, and (f) 998 ms. The horizontal axis shows the time, and the vertical axis shows the output voltage of the detector. The three different colored lines indicate the different pressures of $N_2$ gas at which the experiment was conducted, as shown in the legend.

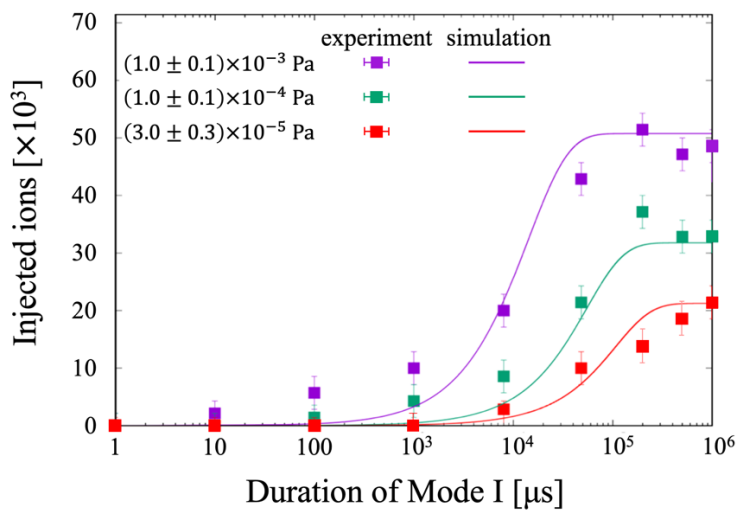

**Figure 10.** Compilation of the results in Figures 8 and 9. The horizontal axis shows the duration of Mode I, and the vertical axis shows the number of injected ions. For the experimental data, points and error bars were calculated from the mean and variance of the maximum amplitude of the detector in three measurements, and the voltage output was calibrated with oscilloscope test signals. For the simulation data, pressure dependence is incorporated in the ion generation rate and in the collision frequency between $N_2^+$ and $N_2$.

### 3.2. Demonstration with a miniature TOF-MS

We subsequently tested the new ion buncher in combination with a miniature TOF-MS[26] to see if this new ionizer could contribute to the sensitivity enhancement. The TOF-MS used in the experiments is a reflectron type and achieves m/Δm > 50 by sending ions back and forth within the optics. The miniature MS originally used a streaming EI ionizer to obtain the pulsed ion beam; here we conducted experimental tests by exchanging the previous ionizer to the newly developed bunching ionizer. Experimentally, the total power consumption of the TOF-MS system with the previous ionizer was ~7.2 W, whereas with the new ion source, it was ~7.8 W. This indicates that the increase in power consumption from the previous system was kept as low as possible.

Figure 11 compares the experimental results with different ionizers. Both measurements were performed in a 99.9% $N_2$ condition at $(1\pm0.1) \times 10^{-4}$ Pa. The major changes in each case are the density and temporal dispersion of pulsed ion beam. For the former, increasing the density of the pulsed ion beam should contribute to an increase in sensitivity, since the expected value of the ion detection becomes higher. For the latter, temporal dispersion leads to an increase in Δm, which should result in a compromise in resolution. As expected, a remarkably positive result was achieved, where the sensitivity to $N_2^+$ increased by more than one order of magnitude. As a side effect, the $N_2^+$ peak broadened and the mass resolution decreased from m/Δm ~ 60 to ~ 40; depending on the target gas species, we consider this compromise in resolution to be acceptable. Interestingly, there was also an increase in the intensity of the *m/z*=14 peak compared to the *m/z*=28 peak. This suggests that the ions appear to have fragmented a certain amount during the bunching time (i.e., $N_2^+ \rightarrow N+N^+$). This demonstrates the potential for contribution to structural analysis of molecules, but also indicates that careful calibration is necessary for quantitative analysis.

Optimization of MS usually encompasses a tradeoff between mass resolution and sensitivity. For example, considering classical MS, one would be able to obtain higher sensitivity by enlarging an entrance slit and/or extending the ion extraction time, which simultaneously reduces the resolution of the mass spectrum. Figure 12 shows the results on attempts to improve sensitivity by the classical methods. Based on the comparision, we report that the results obtained with the new bunching ionizer (Figure 11) showed the superiority in terms of mass resolution.

| | Streaming Ionizer | Bunching Ionizer |
|---|---|---|
| | 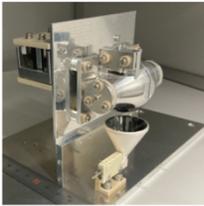 | 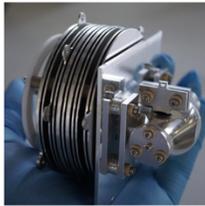 |
| Mass Spectrum Integration: 60 s, Analyte gas: $1\times10^{-4}$ Pa $N_2$ | 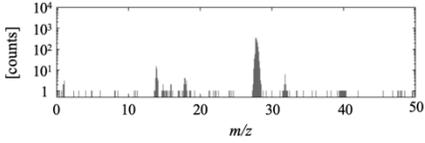 | 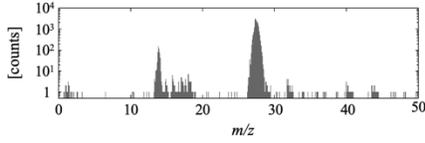 |
| Sensitivity @$m/z$=28 [(counts/s)/Pa] | $1.5\ (\pm 0.1)\times 10^6$ | $2.5\ (\pm 0.1)\times 10^7$ |
| $m/\Delta m$ @$m/z$=28, FWHM | ~60 | ~40 |
| [$m/z$=14]/[$m/z$=28] | $0.020 \pm 0.003$ | $0.092 \pm 0.010$ |

**Figure 11.** Comparison of experimental results obtained with different ionizers. (Left column) Data obtained with a miniature TOF-MS using a streaming ionizer. (Right column) Data obtained with a miniature TOF-MS using the new bunching ionizer. (First raw) Examples of mass spectra obtained in the experiment. Horizontal axes show mass-to-charge ratio, and vertical axes show ion counts in logarithmic scale. Bin widths of spectra are set as $\Delta m/z$=0.010. (Second raw) Sensitivity calculated from the total counts at $m/z$ = 28. The errors are due to statistical errors of counts, which is calculated as the square root of the total counts. (Third raw) Mass resolution estimated at full width half maximum at $m/z$ = 28. (Fourth raw) Relative amount of ion counts for the $m/z$ = 14 peak and the $m/z$ = 28 peak. The errors are estimated from a statistical error (square root of the total counts), background noise level, and range decision corresponding to $m/z$ =14 or 28.

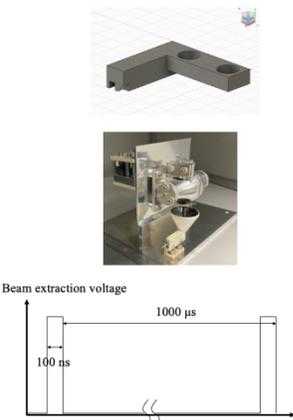

**Figure 12.** Experimental results with a streaming ionizer, attempting to increase sensitivity in classical methods. (First column) Data obtained with a miniature TOF-MS using a streaming ionizer, same as the left column in Figure 11. (Second column) Data obtained with a miniature TOF-MS that doubled the ion extraction time. (Third column) Data obtained with a miniature TOF-MS that enlarged the TOF entrance aperture (1.0×1.0 mm² to 2.5×2.5 mm²). (First raw) Examples of mass spectra obtained in the experiment. Horizontal axes show mass-to-charge ratio, and vertical axes show ion counts in logarithmic scale. Bin widths of spectra are set as Δ$m/z$=0.010. (Second raw) Sensitivity calculated from the total counts at $m/z$ = 28. The errors are due to statistical errors of counts, which is calculated as the square root of the total counts. (Third raw) Mass resolution estimated at full width half maximum at $m/z$ = 28.

## 4. Discussion and Conclusion

In this study, we developed a new bunching ionizer based on the electrostatic ion beam trap principle by revisiting the parameter study of the design. The ionizer combines EI ionization and bunching capabilities and can store ions generated during TOF separation which are not available in streaming ionizers. The new device is small (~25 mm in length) and requires low voltage (~100 V) and power (~2.2 W). In addition, experimental results showed that the ion bunch was obtained as expected in the simulation, and that it contributed to improved sensitivity when tested in combination with a miniature TOF-MS. Another interesting result was that the fraction of fragment ions in the bunched ion beam increased by several times compared with the pulsed ion beam from the streaming ionizer.

These results indicate that the new device could be useful as an ionizer to improve the sensitivity of TOF-MSs, especially for miniature TOF-MSs with reduced resources used in harsh-environments and planetary exploration. However, when connecting this device to other TOF-MSs, the time dispersion of the ion bunch ~1 μs and the energy dispersion of 60–80 eV must be acceptable. This is a major limitation in utilizing this device; hence, it has potential for further development. For example, more confined bunches could be achieved by higher kick-out voltages, post-acceleration of ion bunches, and ejecting ion bunches in a direction perpendicular to the EIBT axis. Besides sensitivity, the increase in multivalent and/or fragment ions indicates its potential use as a fragmentation/reaction cell. This can be studied as future works while considering the balance between resources and performance.

## Acknowledgements

The authors gratefully acknowledge the support for text readability provided by University of Maryland. This work was supported by Kakenhi 23KJ2212, 22K2134.

## Declarations

Conflict of interest: The authors declare no competing interests.

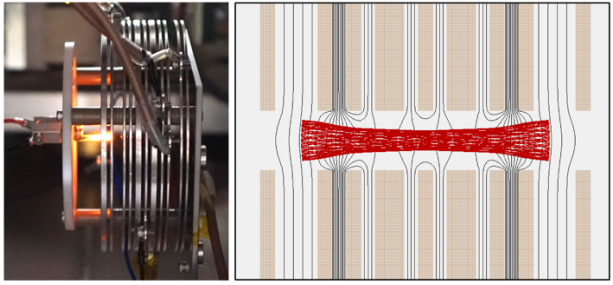

For Table of Contents Use Only

Manuscript title:
Development of a bunching ionizer for TOF mass spectrometers with reduced resources

Names of authors:
Oya Kawashima, Satoshi Kasahara, Yoshifumi Saito, Masafumi Hirahara, Kazushi Asamura, Shoichiro Yokota

A brief synopsis describing the graphic:
In this study, we designed a new electrostatic ion beam trap as a bunching ionizer to improve the sensitivity of the mass spectrometer. This graphic shows the fabricated test model and the typical ion trap trajectory. By combining the test model with a small TOF mass spectrometer, we confirmed that the TOF-MS sensitivity increased by a factor of 10 compared with the use of a conventional ionizer.

Note:
This TOC graphics is original (does not contain copyright-protected material) and created in its entirety by Oya Kawashima, the first author of the manuscript.